\documentclass{article}

\usepackage{amsmath}
\usepackage{PRIMEarxiv}
\usepackage{mathrsfs}
\usepackage{listings}
\usepackage{longtable}
\usepackage{tabularx}
\usepackage{forloop}
\usepackage{etoolbox}
\usepackage{multirow}
\usepackage{subcaption}
\usepackage{float}
\usepackage{calc}
\usepackage{wrapfig,lipsum,booktabs} 
\usepackage{xtab} 
\usepackage[dvipsnames]{xcolor}

\usepackage[utf8]{inputenc} 
\usepackage[T1]{fontenc}    
\usepackage{hyperref}       
\usepackage{url}            
\usepackage{booktabs}       
\usepackage{amsfonts}       
\usepackage{nicefrac}       
\usepackage{microtype}      
\usepackage{lipsum}
\usepackage{fancyhdr}       
\usepackage{graphicx}       
\usepackage{adjustbox}
\graphicspath{{media/}}     

\usepackage{hyperref}
\hypersetup{
    colorlinks=true,       
    linkcolor=blue,        
    citecolor=blue,        
    filecolor=blue,        
    urlcolor=blue          
}

\title{A Framework for Measuring How News Topics Drive Stock Movement}

\author{
  Qizhao Chen \\
  Graduate School of Information Science \\
  University of Hyogo \\
  Kobe, Japan\\
  \texttt{af24o008@guh.u-hyogo.ac.jp} \\
}

\begin{document}
\maketitle

\begin{abstract}
In modern financial markets, news plays a critical role in shaping investor sentiment and influencing stock price movements. However, most existing studies aggregate daily news sentiment into a single score, potentially overlooking important variations in topic content and relevance. This simplification may mask nuanced relationships between specific news themes and market responses. To address this gap, this paper proposes a novel framework to examine how different news topics influence stock price movements. The framework encodes individual news headlines into dense semantic embeddings using a pretrained sentence transformer, then applies K-means clustering to identify distinct news topics. Topic exposures are incorporated as explanatory variables in an ordinary least squares regression to quantify their impact on daily stock returns. Applied to Apple Inc., the framework reveals that certain topics are significantly associated with positive or negative next-day returns, while others have no measurable effect. These findings highlight the importance of topic-level analysis in understanding the relationship between news content and financial markets. The proposed framework provides a scalable approach for both researchers and practitioners to assess the informational value of different news topics and suggests a promising direction for improving predictive models of stock price movement.
\end{abstract}


\section{Introduction}\label{sec1}

Financial markets are highly sensitive to the flow of information, and news has long been recognized as a critical driver of investor sentiment and asset price movements~\cite{10.1145/3694860.3694870,deveikyte2022sentiment}. With the increasing availability of large-scale textual data from financial news outlets, analysts and researchers have turned to natural language processing (NLP) and machine learning techniques to extract signals that can help explain and predict stock returns. A growing body of literature demonstrates that textual features such as sentiment polarity, frequency of coverage, and specific word categories can be correlated with stock market performance~\cite{10825946,math12233801,chen2025dualtransformer}. However, despite these advances, much of the existing work such as~\cite{MAQBOOL20231067,Ouf2024} focuses on aggregate sentiment indicators~\cite{xiao2023stock}, which often overlook the nuanced roles that different kinds of news content may play in shaping market reactions. A positive tone about corporate governance, for example, may have a different economic impact compared to a positive tone about product innovation. This raises an important question: how do distinct news topics, rather than overall sentiment, affect stock price dynamics?

A common limitation in the literature stems from how daily news is aggregated. Each day, there may be numerous articles related to a single company. In many studies~\cite{chen2025sentimentawarestockpriceprediction,srinivas2023effectsdailynewssentiment,chen2025adaptivealphaweightingppo}, the polarity scores of these articles are averaged to obtain a single daily sentiment indicator. While intuitive, this averaging procedure risks discarding valuable information. For instance, a few critical articles about lawsuits or executive turnover may be overshadowed when combined with a larger number of neutral or mildly positive pieces, leading to an overall sentiment score that does not reflect the true market relevance of the news. Simply taking the average may therefore mask the influence of more impactful stories. This motivates the need for an approach that can rank or weight news items by their importance, and assess whether certain types of news systematically drive stock returns more than others.

This paper proposes a simple yet flexible framework to investigate the impact of news topics on stock market movements. The framework consists of three main components. First, news titles are transformed into dense semantic representations using a sentence transformer model, which captures contextual meaning beyond surface-level word matching. Second, the embeddings are grouped into coherent topics through clustering, enabling the identification of recurring themes in the news corpus. Finally, daily topic exposures are constructed and regressed against stock returns to evaluate which types of news coverage are significantly associated with subsequent market performance. This framework moves beyond sentiment-based analysis by incorporating the semantic structure of news and allowing different topics to be assessed individually in terms of their explanatory power for stock returns.

The motivation for this approach is twofold. From a methodological perspective, the use of pretrained transformer models provides a robust and efficient way to capture semantic meaning in short texts such as news headlines, which are often too concise for traditional bag-of-words or sentiment dictionaries to capture effectively. Clustering on the embedding space helps reveal latent themes that may not be apparent at the individual article level but emerge as patterns across the corpus. From a financial perspective, the framework allows us to quantify whether specific topics are systematically related to stock returns. This opens the door to a more granular understanding of how information affects prices and may provide useful signals for investors and risk managers.

The proposed framework contributes to both the finance and NLP literature. For finance, it highlights the importance of distinguishing between different kinds of news rather than treating all coverage as a homogeneous sentiment signal. For NLP, it demonstrates a practical application of sentence embeddings and unsupervised clustering in an empirical finance setting. While the framework is applied here to Apple Inc. (AAPL) as a case study, it can be easily extended to other firms or broader market indices. By aligning daily topic exposures with stock returns, the approach provides interpretable insights into which kinds of information the market responds to most strongly.

\section{Related Work}\label{sec2}

The intersection of textual analysis and financial market prediction has attracted significant attention in recent years. A large body of research has focused on leveraging textual information from financial news and social media to understand and forecast stock price movements. These studies generally fall into three broad categories: sentiment analysis, topic modeling, and combined textual–quantitative frameworks.

The first category, sentiment analysis, examines how positive or negative tone in textual content relates to asset price or return. Traditional approaches rely on dictionaries and lexicons to assign sentiment scores to words or phrases, and then aggregate these scores into document-level or daily-level sentiment measures~\cite{10.1162/COLI_a_00049}. More recent work employs advanced natural language processing techniques, including deep learning models, to extract sentiment features automatically. These sentiment indicators have been used to predict returns, volatility, and trading volume. For example, Saravanos and Kanavos~\cite{10345902} forecast several indices of the U.S. stock market using historical closing prices and sentiment data. Song et al.~\cite{SONG2025108210} also use investor textual sentiment to predict the realized volatility and VaR (Value at Risk).  Alvarez-Diez et al.~\cite{Alvarez-Diez31072025} employ the ChatGPT-extracted sentiment polarity score to predict the intraday abnormal stock return.

Although sentiment-based methods can capture the general mood of the market or specific firms, they often treat all news equally, without differentiating the content or topic of the news items. This can obscure important information embedded in the heterogeneity of news coverage.

The second category focuses on topic modeling, which aims to uncover latent themes in textual data. Methods such as Latent Dirichlet Allocation (LDA)~\cite{10.5555/944919.944937} and non-negative matrix factorization~\cite{Kuang2015} have been applied to financial news and corporate disclosures to extract interpretable topic structures. Topic modeling enables researchers to move beyond sentiment and identify recurring themes such as product announcements, mergers and acquisitions, regulatory developments, or macroeconomic news. Some studies have explored the relationship between topic prevalence and stock movement. For example, Iwasaki et al.~\cite{https://doi.org/10.1111/irfi.12425} integrate the topic tones of the analyst reports to predict cumulative abnormal return. Zhu~\cite{zhu2024bertopicdrivenstockmarketpredictions} uses an NLP model, BERTopic, to analyze the sentiment of topics derived from stock market comments and the results find that topics in stock market comments provide valuable insights into stock market movement.

Another important issue in the literature is how to aggregate news information over time. Many studies compute daily sentiment or topic scores by averaging across all news items for a firm on a given day. While simple, this approach risks overlooking the varying importance of different articles. Not all news items have the same impact on market behavior; certain news may contain more relevant or urgent information, while others may be routine or duplicative~\cite{Wang_2024}. This limitation has motivated recent work to explore weighting schemes, attention mechanisms, or event-based aggregation~\cite{wang2025eventawaresentimentfactorsllmaugmented} to better capture the relative importance of individual news items. However, few studies provide a systematic framework to quantify and rank news items by their impact potential.

With the development of deep learning, embedding models such as Word2Vec~\cite{mikolov2013efficientestimationwordrepresentations}, GloVe~\cite{pennington-etal-2014-glove}, and BERT~\cite{devlin2019bertpretrainingdeepbidirectional} have been applied to represent financial texts in continuous vector spaces, enabling richer semantic analysis. Among these, sentence transformers~\cite{reimers2019sentencebertsentenceembeddingsusing} have gained attention because they extend the BERT architecture with a Siamese or triplet network structure, making them particularly effective for generating sentence-level embeddings. Unlike traditional embeddings that focus on word-level meaning, sentence transformers capture the semantic relationship between entire sentences, allowing for better representation of news headlines and short texts. This makes them well suited for financial applications, where subtle variations in phrasing can imply significantly different market implications. By leveraging these embeddings, researchers can move beyond simple polarity scores to uncover deeper patterns in how different news topics affect stock price movements.

The framework proposed in this paper addresses the gaps by combining sentence transformer embeddings, clustering-based topic modeling, and regression analysis of daily topic exposures. This approach allows for both a semantic understanding of news content and a quantitative assessment of its market impact. Furthermore, it enables the ranking of news articles by their importance to returns, addressing the limitation of simple daily aggregation. In doing so, it contributes to the broader literature by offering a scalable, interpretable, and flexible method for studying the relationship between news topics and stock price movements.

\section{Methodology}

This study proposes a framework to quantify how different news topics affect stock price movements. The methodology consists of three main stages: news embedding generation, topic clustering, and regression analysis. Each stage is described in detail below.

\subsection{Data}

The Apple stock price data from 2016 to 2024 (Table~\ref{tab:apple_stock_sample}) are downloaded from Yahoo Finance using the yfinance Python library. Daily returns are calculated as the percentage change in closing prices. News data of Apple are downloaded using an API\footnote{https://eodhd.com/financial-apis/stock-etfs-fundamental-data-feeds}. Table~\ref{tab:news_sample} shows a sample of the news dataset used in this paper.

\begin{table*}[htbp]
\centering
\caption{Sample of Apple Daily Stock Prices}
\begin{tabular}{lccccc}
\hline
\textbf{Date} & \textbf{Open} & \textbf{High} & \textbf{Low} & \textbf{Close} & \textbf{Volume} \\
\hline
2016-02-19 & 21.783560 & 21.956014 & 21.738178 & 21.792637 & 141496800 \\
2016-02-22 & 21.853912 & 21.987791 & 21.765416 & 21.983252 & 137123200 \\
2016-02-23 & 21.874336 & 21.897027 & 21.454549 & 21.486317 & 127770400 \\
2016-02-24 & 21.325205 & 21.869792 & 21.175442 & 21.806257 & 145022800 \\
2016-02-25 & 21.794913 & 21.956020 & 21.613383 & 21.956020 & 110330800 \\
... & ... & ... & ... & ... & ... \\
2024-05-01 & 168.370898 & 171.478586 & 167.904248 & 168.092896 & 50383100 \\
2024-05-02 & 171.279999 & 172.183514 & 169.671554 & 171.796295 & 94214900 \\
2024-05-03 & 185.319184 & 185.666694 & 181.357642 & 182.072510 & 163224100 \\
2024-05-06 & 181.049850 & 182.886650 & 179.133603 & 180.414413 & 78569700 \\
2024-05-07 & 182.142004 & 183.581662 & 180.027201 & 181.099487 & 77305800 \\
\hline
\end{tabular}
\label{tab:apple_stock_sample}
\end{table*}

\begin{table*}[htbp]
\centering
\caption{Sample of Apple News Headlines Dataset}
\begin{tabular}{clp{9cm}}
\hline
\textbf{Index} & \textbf{Date} & \textbf{Title} \\
\hline
0 & 2024-05-08 & Judge grills Apple exec about whether company \dots \\
1 & 2024-05-08 & Apple’s Unionized Maryland Store to Vote on Po\dots \\
2 & 2024-05-08 & Music streaming firms urge European Commission\dots \\
3 & 2024-05-08 & The 3 Best Metaverse Stocks to Buy in May 2024 \\
4 & 2024-05-08 & Apple’s iPad event was an AI teaser for its fu\dots \\
... & ... & ... \\
26745 & 2018-01-31 & Investor Expectations to Drive Momentum within\dots \\
26746 & 2017-11-30 & BioTelemetry, Inc. Enters Agreement to Provide\dots \\
26747 & 2017-11-27 & Factors of Influence in 2018, Key Indicators a\dots \\
26748 & 2017-10-05 & New Research: Key Drivers of Growth for Micros\dots \\
26749 & 2016-02-19 & Payment Data Systems Announces Apple Pay Suppo\dots \\
\hline
\end{tabular}
\label{tab:news_sample}
\end{table*}

\subsection{News Embedding Generation}

To capture the semantic meaning of each news headline, a sentence transformer model is used. Specifically, the pretrained \texttt{all-MiniLM-L6-v2} model is adopted to transform each headline into a dense vector representation in a high-dimensional embedding space. This approach enables capturing contextual and semantic information that goes beyond surface-level word matching.

A sentence transformer is a type of neural network model designed to map sentences or short texts into fixed-size dense vectors, called embeddings. These embeddings encode the semantic meaning of the input text, allowing similarity comparisons and downstream tasks such as clustering or classification to be performed efficiently.

Let \( H_t^i \) denote the headline of article \( i \) on day \( t \), and \( E(H_t^i) \) the embedding vector generated by the model. These embeddings form the input for the clustering stage.

\subsection{Topic Clustering}

K-means clustering is applied to the headline embeddings to group semantically similar headlines into distinct topics. Instead of arbitrarily fixing the number of topics \( K \), the optimal number of clusters is determined using the silhouette score. The silhouette score measures how well each headline fits within its assigned cluster compared to other clusters, with values ranging from \(-1\) to \(1\). A higher score indicates that the clusters are well-separated and internally cohesive. Based on this evaluation, the number of topics is set to \( K=5 \), which strikes a balance between interpretability and clustering quality. Each headline is then assigned a topic label \( k \in \{0, 1, \dots, K-1\} \) according to the nearest cluster centroid in the embedding space. These topic labels are appended to the dataset for further analysis.

K-means clustering is an unsupervised machine learning algorithm that partitions data into \( K \) clusters by minimizing the within-cluster variance. Each cluster is represented by its centroid, and data points are assigned to the cluster with the nearest centroid. This method is computationally efficient and widely used for topic discovery and grouping of textual embeddings.

\begin{figure}[!htbp]
    \centering
    \includegraphics[width=0.7\textwidth]{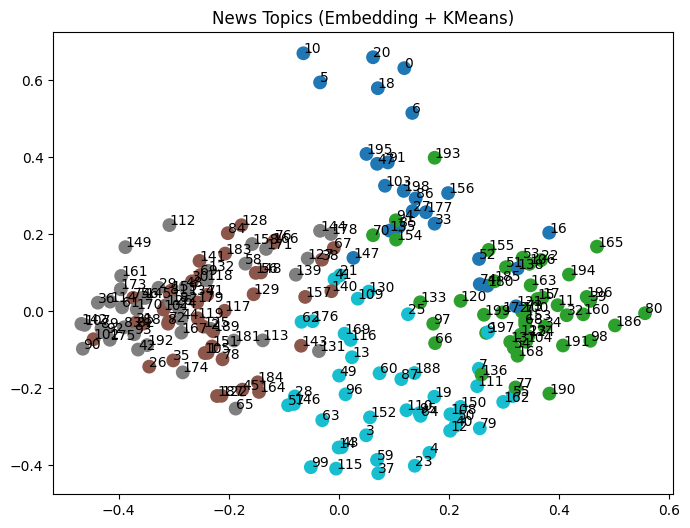}  
    \caption{Topic Clustering Example}
    \label{fig: Topic Clustering Example}
\end{figure}

To visualize the clustering results, Principal Component Analysis (PCA) is applied to reduce the high-dimensional embeddings to two dimensions. This facilitates plotting and interpretation of topic clusters. Figure~\ref{fig: Topic Clustering Example} demonstrates a topic clustering example.

\subsection{Regression Analysis}

To measure the relationship between topics and stock returns, an ordinary least squares (OLS) regression is performed. 

News data and stock returns are merged by date. Topic labels are transformed into dummy variables representing the presence of each topic on a given day. The regression model is specified as:

\[
R_t = \alpha + \sum_{k=0}^{K-1} \beta_k D_{t,k} + \epsilon_t
\]

Where:
\begin{itemize}
    \item \(R_t\) is the stock return for day \(t\),
    \item \(D_{t,k}\) is a binary dummy variable indicating whether topic \(k\) is present on day \(t\),
    \item \(\beta_k\) captures the effect of topic \(k\) on the return,
    \item \(\alpha\) is the intercept, and \(\epsilon_t\) is the error term.
\end{itemize}

Dummy variables are generated using one-hot encoding for each topic, and a constant term is added to the regression model to capture the intercept. Missing return values are replaced with zero to ensure alignment with the news dataset.

The regression results allow identifying which topics are significantly associated with stock returns and estimating the magnitude of their impact.

\begin{table*}[!htbp]
\centering
\caption{OLS regression results for daily stock return of Apple Inc. using topic dummies.}
\label{tab:ols_results}
\begin{tabular}{lcccccc}
\hline
Variable & Coefficient & Std. Error & t-value & P-value & 95\% CI Lower & 95\% CI Upper \\
\hline
const    & -0.0010     & 0.001       & -1.012  & 0.313   & -0.003         & 0.001          \\
Topic\_0 & -0.0058     & 0.003       & -2.112  & 0.036   & -0.011         & -0.000         \\
Topic\_1 & 0.0004      & 0.002       & 0.210   & 0.834   & -0.004         & 0.005          \\
Topic\_2 & -0.0003     & 0.002       & -0.124  & 0.902   & -0.005         & 0.004          \\
Topic\_3 & 0.0001      & 0.002       & 0.054   & 0.957   & -0.004         & 0.004          \\
Topic\_4 & 0.0045      & 0.002       & 2.097   & 0.037   & 0.000          & 0.009          \\
\hline
\end{tabular}
\end{table*}

\begin{table*}[!htbp]
\centering
\caption{Topic importance ranking for Apple Inc. news topics.}
\label{tab:topic_importance}
\begin{tabular}{lc}
\hline
\textbf{Topic} & \textbf{Importance Score} \\
\hline
Topic\_0 & 0.519 \\
Topic\_4 & 0.406 \\
Topic\_1 & 0.040 \\
Topic\_2 & 0.024 \\
Topic\_3 & 0.011 \\
\hline
\end{tabular}
\end{table*}

\subsection{Topic Importance Score}

After estimating the regression model, we assess the relative importance of each news topic in explaining stock return variation. 
The regression coefficients associated with each topic dummy represent the marginal effect of topic exposure on daily returns. 
However, to make the effects comparable and interpretable, we compute a normalized measure of \textit{topic importance}.  

Specifically, we first take the absolute value of each coefficient (excluding the constant term) to capture the magnitude of impact regardless of direction (positive or negative). 
Then, these values are normalized by dividing by their total sum, ensuring that the importance scores across topics add up to one. 
This normalization provides a clear ranking of which topics exert stronger influence on stock movements.  

Formally, if $\hat{\beta}_k$ is the estimated coefficient for topic $k$, then the normalized topic importance $I_k$ is defined as:  

\begin{equation}
I_k = \frac{|\hat{\beta}_k|}{\sum_{j=1}^{K} |\hat{\beta}_j|}
\end{equation}

This ranking highlights the most influential news themes, allowing us to identify which types of information investors appear to respond to most strongly in the market.

\section{Results}

The regression analysis investigates the relationship between news topic exposures and the daily stock return of Apple Inc. The model regresses the next-day return on topic dummies derived from clustering news headlines, with the constant term capturing the baseline return. The results are summarized in Table~\ref{tab:ols_results}.

The model achieves an \(R^2\) of 0.035, suggesting that topic exposures explain a small but non-negligible portion of the variation in daily returns. The overall F-statistic is 1.745 with a p-value of 0.142, indicating that the model as a whole is marginally insignificant at conventional significance levels.

Among the five topics, Topic\_0 and Topic\_4 have statistically significant coefficients at the 5\% level. Topic\_0 exhibits a negative coefficient of -0.0058 (\(p = 0.036\)), indicating that days with greater exposure to this topic are associated with lower next-day returns. Conversely, Topic\_4 has a positive coefficient of 0.0045 (\(p = 0.037\)), suggesting that higher exposure to this topic is associated with positive stock returns. The other topics (Topic\_1, Topic\_2, Topic\_3) do not show statistically significant effects.

The residual diagnostics show a Durbin–Watson statistic of 1.932, indicating no serious autocorrelation. However, the Omnibus and Jarque–Bera tests indicate departures from normality in the residuals (\(p < 0.01\)), and the skewness and kurtosis values suggest slight asymmetry and heavier tails than a normal distribution. These issues suggest that further robustness checks and potentially nonlinear models could improve the analysis.

Table~\ref{tab:topic_importance} shows the topic importance scores for all five topics. Based on the scores, Topic\_0 and Topic\_4 have a larger impact on stock returns compared to the others.

Overall, these findings demonstrate that not all news topics carry equal influence over stock returns. The statistically significant impact of certain topics highlights the importance of topic-level analysis when examining the relationship between textual news data and financial performance. This suggests that investors and researchers could gain additional insight by differentiating news content rather than relying solely on aggregated sentiment measures.

\section{Further Discussion}

Although this study primarily employed Sentence Transformer embeddings combined with K-Means clustering to identify semantic patterns in financial news, one limitation of this approach is the difficulty in interpreting the resulting topics. While K-Means can effectively group similar headlines based on their embeddings, it does not provide explicit information about what each cluster represents, making topic interpretation largely subjective. In contrast, Latent Dirichlet Allocation (LDA) and BERTopic~\cite{grootendorst2022bertopicneuraltopicmodeling} offer more interpretable outputs. LDA, as a probabilistic generative model, assigns words to latent topics, enabling researchers to examine the most representative keywords for each topic. Similarly, BERTopic builds on transformer-based embeddings and applies class-based TF–IDF to extract meaningful keywords for each cluster, providing clearer insights into the underlying semantic themes. Therefore, both LDA and BERTopic can enhance the interpretability of topic modeling results by explicitly revealing the dominant words associated with each topic, which helps in understanding how different news narratives may influence stock movements.

Table~\ref{tab:lda_keywords} presents the top ten keywords for each topic identified using the LDA model. These keywords provide insight into the main themes represented in the news headlines. For example, Topic 0 appears to capture discussions related to technology companies and product updates, featuring terms such as “apple,” “google,” “amazon,” and “microsoft.” Topic 1 focuses more on financial and market-related contexts, including words like “dow,” “futures,” and “fed,” suggesting coverage of stock indices and macroeconomic events. Topic 2 contains investment-related terms such as “buy,” “portfolio,” and “dividend,” indicating articles emphasizing trading and portfolio strategies. Topic 3 revolves around company fundamentals and performance indicators, with frequent mentions of “earnings,” “sales,” and “billion.” Finally, Topic 4 combines general stock market dynamics and investor sentiment, featuring words such as “stocks,” “market,” and “investors.” Overall, the extracted keywords show that LDA can uncover distinct yet interrelated themes within financial news, which can then be used to analyze how different types of information influence stock movements.

\begin{table*}[htbp]
\centering
\caption{Top 10 Keywords for Each Topic Identified by LDA}
\label{tab:lda_keywords}
\begin{tabular}{cl}
\hline
\textbf{Topic} & \textbf{Top Keywords} \\
\hline
Topic 0 & apple, new, google, tech, amazon, microsoft, watch, update, big, says \\
Topic 1 & apple, dow, jones, iphone, futures, market, tesla, amp, app, fed \\
Topic 2 & stocks, buy, 10, best, buffett, warren, apple, dividend, portfolio, according \\
Topic 3 & apple, earnings, sales, stock, iphone, 2024, billion, china, markets, shares \\
Topic 4 & stocks, stock, apple, buy, market, tech, today, big, aapl, investors \\
\hline
\end{tabular}
\end{table*}

Table \ref{tab:bertopic_results} presents a summary of the BERTopic analysis, listing the main topics identified in the dataset. For each topic, the table shows its identifier, the number of documents associated with it, a descriptive name, and representative keywords and sample documents. Topic $-1$, for example, reflects broad market trends with keywords such as “jones,” “dow,” and “nasdaq,” while Topic 0 focuses on growth stocks, and Topic 1 centers on iPhone demand. This table provides a clear overview of the dominant themes and their prevalence, offering insight into the underlying structure of the analyzed text data.

\begin{table*}[htbp]
\centering
\caption{Summary of BERTopic Results}
\label{tab:bertopic_results}
\adjustbox{max width=\textwidth}{
\begin{tabular}{ccll}
\hline
\textbf{Topic} & \textbf{Count} & \textbf{Name} & \textbf{Representative Keywords / Documents} \\
\hline
-1 & 7656 & -1\_jones\_dow\_on\_nasdaq & [jones, dow, on, nasdaq, futures, stocks, stoc...] \\ 
   &      &                           & [Dow Jones Futures Rise Ahead Of Key Economic ...] \\
0 & 519 & 0\_growth\_stocks\_hold\_forever & [growth, stocks, hold, forever, richer, that, ...] \\ 
   &     &                                 & [Want to Get Richer? 2 Top Stocks to Buy Now a...] \\
1 & 376 & 1\_iphone\_14\_15\_demand & [iphone, 14, 15, demand, production, pro, 13, ...] \\ 
   &     &                           & [How the iPhone 14, 14 Plus and 14 Pro Compare...] \\
2 & 345 & 2\_big\_tech\_techs\_season & [big, tech, techs, season, earnings, are, this...] \\ 
   &     &                            & [Big Techs' Growing Finance Business Is Gettin...] \\
3 & 325 & 3\_sampp\_500\_slips\_flat & [sampp, 500, slips, flat, as, fed, climbs, ene...] \\ 
   &     &                           & [US STOCKS-S\&P 500 flat as Tesla offsets l...] \\
\vdots & \vdots & \vdots & \vdots \\
426 & 10 & 426\_disney\_farr\_prized\_gloomy & [disney, farr, prized, gloomy, snatch, taketwo...] \\ 
    &    &                                  & [Better Buy: Apple Stock vs. Disney Stock, Bet...] \\
427 & 10 & 427\_valuable\_worlds\_company\_longer & [valuable, worlds, company, longer, title, imm...] \\ 
    &    &                                        & [Apple Is No Longer the World's Most Valuable ...] \\
428 & 10 & 428\_parental\_child\_parents\_controls & [parental, child, parents, controls, kids, mis...] \\ 
    &    &                                        & [Apple Admits to Bug in Screen Time Parental C...] \\
429 & 10 & 429\_gained\_has\_billion\_buffett & [gained, has, billion, buffett, 181, 158, 182,...] \\ 
    &    &                                     & [Warren Buffett Has Gained Over \$171 Billion O...] \\
430 & 10 & 430\_yields\_treasury\_slashes\_skids & [yields, treasury, slashes, skids, flatten, ju...] \\ 
    &    &                                       & [Dow Slashes Gains, Tech Stocks Slide As Yield...] \\
\hline
\end{tabular}
}
\end{table*}

\begin{figure}[!htbp]
    \centering
    \includegraphics[width=0.7\textwidth]{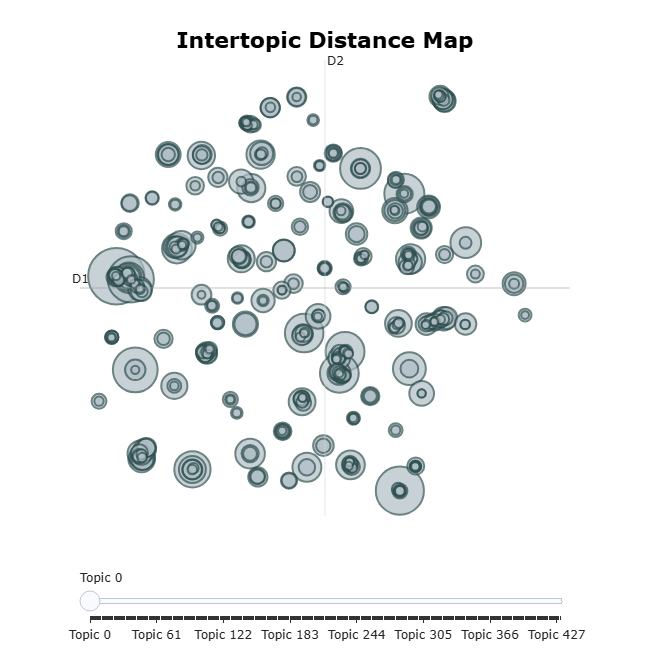}  
    \caption{Intertopic Distance Map Generated by BERTopic}
    \label{fig: intertopic distance map}
\end{figure}

Figure~\ref{fig: intertopic distance map} presents the intertopic distance map generated by BERTopic, which visualizes the semantic relationships among the identified topics. Each circle represents a distinct topic, with its size proportional to the number of documents assigned to it. The positions of the topics are derived from a two-dimensional UMAP projection of their high-dimensional embeddings, such that topics located in closer proximity exhibit greater semantic similarity, while those farther apart represent more dissimilar thematic content. This visualization facilitates the identification of topic clusters reflecting related themes, as well as isolated topics that capture unique or niche subjects within the corpus.

\section{Conclusion}\label{sec5}

This study proposes a framework to examine how different news topics impact stock price movements by combining sentence embeddings, clustering, and regression analysis. Using a pretrained sentence transformer, news headlines are converted into dense vector representations that capture their semantic meaning. K-means clustering is then applied to group these headlines into distinct topics, and the relationship between topic exposures and daily stock returns is quantified through ordinary least squares regression.

The results show that not all topics carry equal weight in influencing stock returns. Specifically, certain topics are significantly associated with positive or negative returns, suggesting that topic-level analysis provides deeper insight than aggregated sentiment measures. These findings underscore the importance of differentiating news content when examining the impact of news on financial markets.

While this study provides a foundational framework, there are several avenues for future research. First, the current approach uses a simple one-hot encoding of topic presence. Future studies could explore dynamic topic modeling methods or probabilistic topic assignments to capture evolving market narratives. Second, expanding the framework to incorporate additional textual features, such as sentiment scores or news source credibility, could improve explanatory power. Third, applying more advanced regression or machine learning models may better capture nonlinear relationships between news topics and stock returns. Finally, extending the framework across multiple companies and markets would allow for testing its robustness and generalizability.

Overall, this framework offers a promising approach to quantify the relationship between textual news content and stock market movements, providing both researchers and practitioners with a tool to better understand how the market interprets and reacts to different types of news.

\bibliographystyle{unsrt}  
\bibliography{references}

\end{document}